\documentclass[aps,prl,twocolumn,showpacs,superscriptaddress,groupedaddress]{revtex4}
\usepackage{graphicx}
\usepackage{dcolumn} 
\usepackage{bm}    
\usepackage{amssymb}  
\usepackage{amsmath}

\renewcommand\th{\vartheta}
\renewcommand{\deg}{^{\circ}}
\newcommand\kip{K_{ip}}

\begin{document}

\title{Chemical Enhancements in Shock-accelerated Particles: Ab-initio Simulations}

\author{Damiano Caprioli}

\affiliation{Department of Astronomy and Astrophysics, University of Chicago, Chicago, IL 60637, USA}
\affiliation{Department of Astrophysical Sciences, Princeton University, Princeton, NJ 08544, USA}
\author{Dennis T.\ Yi}
\affiliation{Department of Astrophysical Sciences, Princeton University, Princeton, NJ 08544, USA}
\author{Anatoly Spitkovsky}
\affiliation{Department of Astrophysical Sciences, Princeton University, Princeton, NJ 08544, USA}

\date{\today}

\begin{abstract}
We study the thermalization, injection, and acceleration of ions with different mass/charge ratios, $A/Z$, in non-relativistic collisionless shocks via hybrid (kinetic ions--fluid electrons) simulations.
In general, ions thermalize to a post-shock temperature proportional to $A$. 
When diffusive shock acceleration is efficient, ions develop a non-thermal tail whose extent scales with $Z$ and whose normalization is enhanced as $(A/Z)^2$, so that incompletely-ionized heavy ions are preferentially accelerated.
We discuss how these findings can explain observed heavy-ion enhancements in Galactic cosmic rays.
\end{abstract}

\maketitle

\emph{Introduction.---} 
Non-relativistic shocks are well-known as sources of energetic particles. 
Prominent examples of such shocks are the blast waves of supernova remnants (SNRs), which are thought to be the sources of Galactic cosmic rays (GCRs) \citep[e.g.,][]{crspectrum,tycho}, and heliospheric shocks, where solar energetic particles (SEPs) are measured in situ \citep[e.g.,][]{reames15,desai+16a}. 
Chemical abundances in GCRs and SEPs provide crucial information about their sources and the processes responsible for their acceleration.

At trans-relativistic energies, the chemical composition of GCRs roughly resembles the composition of the solar system \citep{simpson83}, the most evident deviation being the enhancement of secondaries produced by spallation of primary GCRs during their propagation in the Milky Way.
A more careful analysis, however, reveals that the GCR composition is controlled by volatility and mass/charge ratios: refractory elements show larger enhancements than volatile ones, and heavier volatile elements are more abundant than lighter ones \cite{mde97,edm97}.
Moreover, elemetns with low first ionization potential tend to be overrepresented in GCRs \citep[e.g.,][]{simpson83}.
At TeV energies, where spallation is negligible, the fluxes of H, He, C-N-O, and Fe do not differ by more than one order of magnitude \citep[e.g.,][and references therein]{nuclei}.
Since their typical solar number abundances relative to H are $\chi_{He}=0.0963$, $\chi_{CNO}=9.54\times 10^{-4}$, $\chi_{Fe}=8.31\times 10^{-5}$ \citep{lodders03}, the abundances observed in GCRs suggest that heavy ions must be preferentially injected and accelerated compared to protons.
 
Diffusive shock acceleration (DSA) \citep[e.g.,][]{bell78a,bo78} at SNR shocks is likely the mechanism responsible for ion acceleration up to $\sim 10^{17}$ eV \citep{icrc15}.
DSA produces universal power-law momentum spectra $f(p)\propto p^{-3r/(r-1)}$, where $r$ is the shock compression ratio; for strong shocks $r\to 4$ and $f(p)\propto p^{-4}$.
For relativistic particles the energy spectrum is then $f(E)=4\pi p^2 f(p)dp/dE\propto E^{-2}$, while at non-relativistic energies one gets $f(E)\propto E^{-3/2}$ \citep{DSA}.

%%%%%%%%%%%%%%%%%%%%%%%%%%%%%
\emph{Hybrid simulations.---} 
In order to study ab initio how DSA of ions with different mass/charge ratio works, we performed 2D kinetic simulations with \emph{dHybrid}, a massively parallel hybrid code, in which ions are treated kinetically and electrons as a neutralizing fluid  \citep{gargate+07}.
Hybrid simulations of non-relativistic shocks have been extensively used for assessing the efficiency of proton DSA \citep{DSA}, the generation of magnetic turbulence due to plasma instabilities driven by accelerated particles \citep{MFA}, the diffusion of energetic particles in such self-generated magnetic fields \citep{diffusion}, and the injection of protons into the DSA process \cite{injection}.
In the literature there are few examples of kinetic simulations with heavy ions, e.g., the pioneering 1D hybrid simulations of weak shocks including $\alpha-$particles \citep{burgess89,ts91} and the recent hybrid study of the thermalization of weakly-charged ions at shocks \cite{kropotina+16}.
However, a self-consistent kinetic characterization of ion enhancement in DSA has never been performed before \citep[see, e.g,][for Monte Carlo and semi-analytical approaches]{emp90,nuclei}.
\begin{figure*}
\begin{center}
\includegraphics[trim=0px 50px 0px 340px, clip, width=0.8\textwidth]{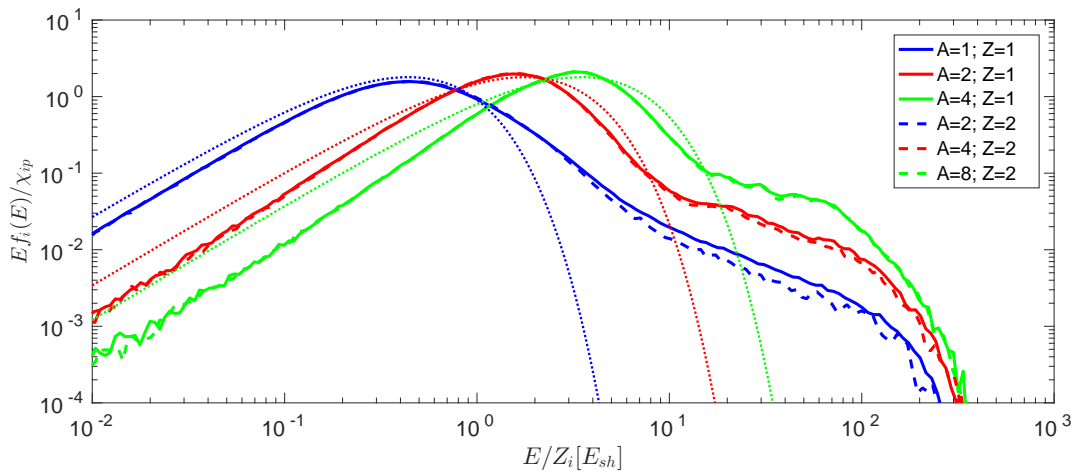}
\caption{Normalized post-shock spectra for ion species with mass $A$ and charge $Z$ as in the legend, for a quasi-parallel ($\th=20\deg$) shock with $M=10$.
The thermal peaks correspond to the Maxwellian distributions (color-matching dotted lines) expected if the temperature scaled with $A$ (see text for more details); the non-thermal tails have a maximum extent $\propto E/Z$ and a normalization  enhanced as a function of $A/Z$. }
\label{fig:spec}
\end{center}
\end{figure*}

In the hybrid simulations presented in this Letter  we include additional ion species characterized by number abundances $\chi_i$, atomic mass $A_i$, and charge $Z_i$ (in proton units), initially in thermal equilibrium with protons and electrons.
We fix $\chi_{i\neq H}=10^{-5}$ to effectively make ions other than protons dynamically unimportant.
Lengths are measured in units of $c/\omega_p$, where $c$ is the speed of light and $\omega_p\equiv \sqrt{4\pi n e^2/m}$, with $m,e$ and $n$ the proton mass, charge and number density;
time is measured in units of $\omega_c^{-1}\equiv mc/eB_0$, $B_0$ being the strength of the initial magnetic field;
velocities are normalized to the Alfv\'en speed $v_A\equiv B/\sqrt{4\pi m n}$, and energies to $E_{sh}\equiv mv_{sh}^2/2$, with $v_{sh}$ the velocity of the upstream fluid in the downstream frame.
We account for the three spatial components of the particle momentum and of the electric and magnetic fields. 
Shocks are produced by sending a supersonic flow against a reflecting wall and are characterized by their sonic and Alfv\'enic Mach numbers $M_s\equiv v_{sh}/c_s$, $M_A\equiv v_{sh}/v_A$, with $c_s$ the sound speed; 
in this work we consider $M_s\simeq M_A\equiv M$.
The shock inclination is defined by the angle $\th$ between the direction of ${\bf B}_0$ and the shock normal, such that $\th\lesssim 45\deg$ corresponds to quasi-parallel shocks.

The time-step is chosen as $\Delta t=0.01/M \omega_c^{-1}$ and the computational box measures $2.5\times 10^4c/\omega_p$ by $2Mc/\omega_p$, with two cells per ion skin depth.
In order to suppress the numerical heating that can arise in long-term simulations with species of disparate densities, we use 100 protons per cell, and 4 particles per cell for all the other species.
We have checked the convergence of our results against 3D simulations, time and space resolution, number of particles per cell, and transverse size of the simulation box \citep[see also][]{DSA}.
The electron pressure is a polytrope with an effective adiabatic index chosen to satisfy the shock jump conditions with thermal equilibration between downstream protons and electrons \citep{injection}.

Our benchmark case comprises ion species with $A=\{1,2,4,8\}$ and $Z=\{1,2\}$ and a quasi-parallel ($\th=20\deg$) shock with $M=10$, which exhibits efficient proton DSA and magnetic field amplification \citep{DSA,MFA}. In our case we find that $\sim 10\%$ of the shock kinetic energy is converted into accelerated protons, and the field is amplified by a factor of $\gtrsim 2$ in the upstream.
%We discuss the role of shock strength and inclination in the next sections.
The downstream spectra of different ion species are shown in Fig.~\ref{fig:spec}, as a function of $E/Z$ and normalized to their abundances $\chi_i$. 
The color code gathers species with the same $A/Z$, while solid and dashed lines correspond to  $Z=1$ and 2, respectively.
Each of the species shows a thermal peak plus a power-law tail with the universal DSA slope $\gamma\simeq 3/2$; non-thermal spectra roll over at a maximum energy $E_{max,i}$, which increases linearly with time \citep{diffusion}.
For strong shocks, Rankine--Hugoniot conditions return a downstream thermal energy $\mathcal E\simeq 0.6E_{sh}$ \citep{DSA}. 
Since half of the post-shock proton energy goes into electron heating by construction, we expect $\mathcal E_{H}\simeq \mathcal E/2$.
Then, since heavier ions have more kinetic energy to convert into thermal energy, their temperature is expected to scale with their masses, i.e.,  $\mathcal E_{i\neq H}=A_i\mathcal E$. 
Dotted lines in Fig.~\ref{fig:spec} correspond to Maxwellian distributions with such expected temperatures: they provide a good fit for the positions of thermal peaks, but only a rough one for the shape of the thermal distributions of heavy ions, whose relaxation is still ongoing \footnote{We checked that the scaling $\mathcal E_i\propto A_i$ is recovered also for protons if the electron pressure is set to zero.}. 

When comparing different ion curves in Fig.~\ref{fig:spec}, we notice three important scalings:
\begin{enumerate}
\item At fixed $Z$, the thermal peaks are shifted to the right linearly in $A$, i.e, each species thermalizes at a temperature proportional to its mass \citep[see also][]{kropotina+16};
\item All the ion spectra rollover at the same $E_{max}/Z$, consistent with the fact that DSA is a rigidity-dependent process \footnote{Rigidity is defined as $p/Z$, not $E/Z$: the two definitions are equivalent only for relativistic particles. However, we showed in \citep{diffusion} that the self-generated diffusion coefficient and, in turn, the acceleration time do scale as $E/Z$ in the non-relativistic case \cite[see also][]{er85}.};
\item The normalization of the non-thermal spectra at given $E/Z$ is an increasing function of the mass/charge ratio, which implies that the efficiency of injection into DSA depends on $A/Z$.
\end{enumerate}
The first two results validate the theoretical expectations, while the last one represents the first self-consistent characterization of the parameter that regulates the injection of ions into the DSA process.

%%%%%%%%%%%%%%%%%%%%%%%%%%%%%
\emph{Injection enhancement in DSA.---} In this section we discuss how the observed boost in ion injection depends on $A/Z$.
The ion non-thermal spectra, neglecting the cutoffs, are power laws that can be written as 
\begin{equation}
f_i(E)=\frac{(\gamma-1) n\chi_i\eta_i }{E_{inj,i}} \left(\frac{E}{E_{inj,i}}\right)^{-\gamma},
\end{equation}
where $\eta_i$ is the fraction of ions that enter DSA above the injection energy $E_{inj,i}$. 
We then introduce the ratio
\begin{equation}\label{eq:kip}
\kip\equiv \frac{f_i(E/Z_i)}{\chi_i f_p(E)}= \frac{\eta_i}{\chi_i\eta_p}\left(\frac{E_{inj,i}}{E_{inj,p}}\right)^{\gamma-1}
\end{equation}
as a measure of the enhancement in energetic ions with respect to protons at fixed $E/Z$. 
$\kip$ is promptly read from Fig.~\ref{fig:spec} by taking the ratio of the power-law spectra at any $E/Z$ between 10 and 100$E_{sh}$. 
Note that the enhancement has two contributions: one straightforward, $\eta_i/\eta_p$, which depends on the fraction of particles that enter DSA for each species, and one more subtle that depends on $E_{inj,i}$, which cannot be predicted analytically.

\begin{figure}
\begin{center}
\includegraphics[trim=0px 20px 10px 227px, clip=true, width=0.49\textwidth]{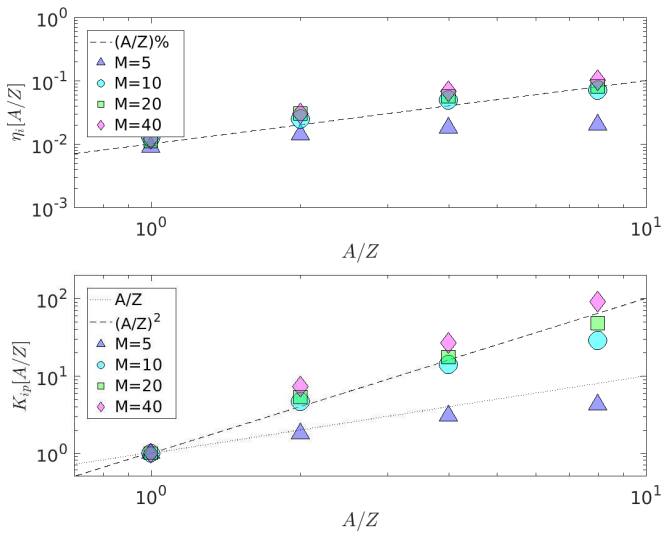}
\includegraphics[trim=0px 225px 10px 20px, clip=true, width=0.49\textwidth]{eta.eps}
\caption{Preferential acceleration of ions with large $A/Z>1$, at $t\approx 10^3\omega_c^{-1}$ for quasi-parallel shocks with Mach numbers as in the legends.
For shocks with $M\gtrsim 10$, the fraction of injected ions $\eta_i$ is linear in $A/Z$ (top panel), while the ion enhancement defined in Eq.~\ref{eq:kip} scales as $(A/Z)^2$ (bottom panel).
For the $M=5$ shock, where self-generated magnetic turbulence is significantly weaker, ion over-injection with respect to protons is less effective, with $K_{ip}$ going roughly as $A/Z$.}
\label{fig:eta}
\end{center}
\end{figure}

Fig.~\ref{fig:eta} shows the enhancements obtained for shocks with $\th=20$ and $M=\{5,10,20,40\}$; 
injection fractions and enhancements are calculated at time $t=10^3\omega_c^{-1}$, when DSA spectra have been established, by considering the post-shock spectra of species with $A/Z$ up to 8, integrated over $10^3 c/\omega_p$.

For shocks with $M\gtrsim 10$, where accelerated protons generate non-linear upstream magnetic turbulence with $\delta B/B_0\gtrsim 1$, the fraction of injected particles is $\eta_p\approx 1\%$ for protons and increases linearly with $A/Z$ (top panel);
at the same time, $\kip\propto (A/Z)^2$, attesting to a very effective enhancement of particles with large charge/mass (bottom panel).
The scaling with $A/Z$ is weaker for the lowest-$M$ shock, for which $\delta B/B_0\approx 0.2$: $\eta_i$ is roughly constant at the percent level and $\kip\propto A/Z$. 

\begin{figure}
\begin{center}
\includegraphics[trim=0px 25px 20px 230px, clip=true, width=0.49\textwidth]{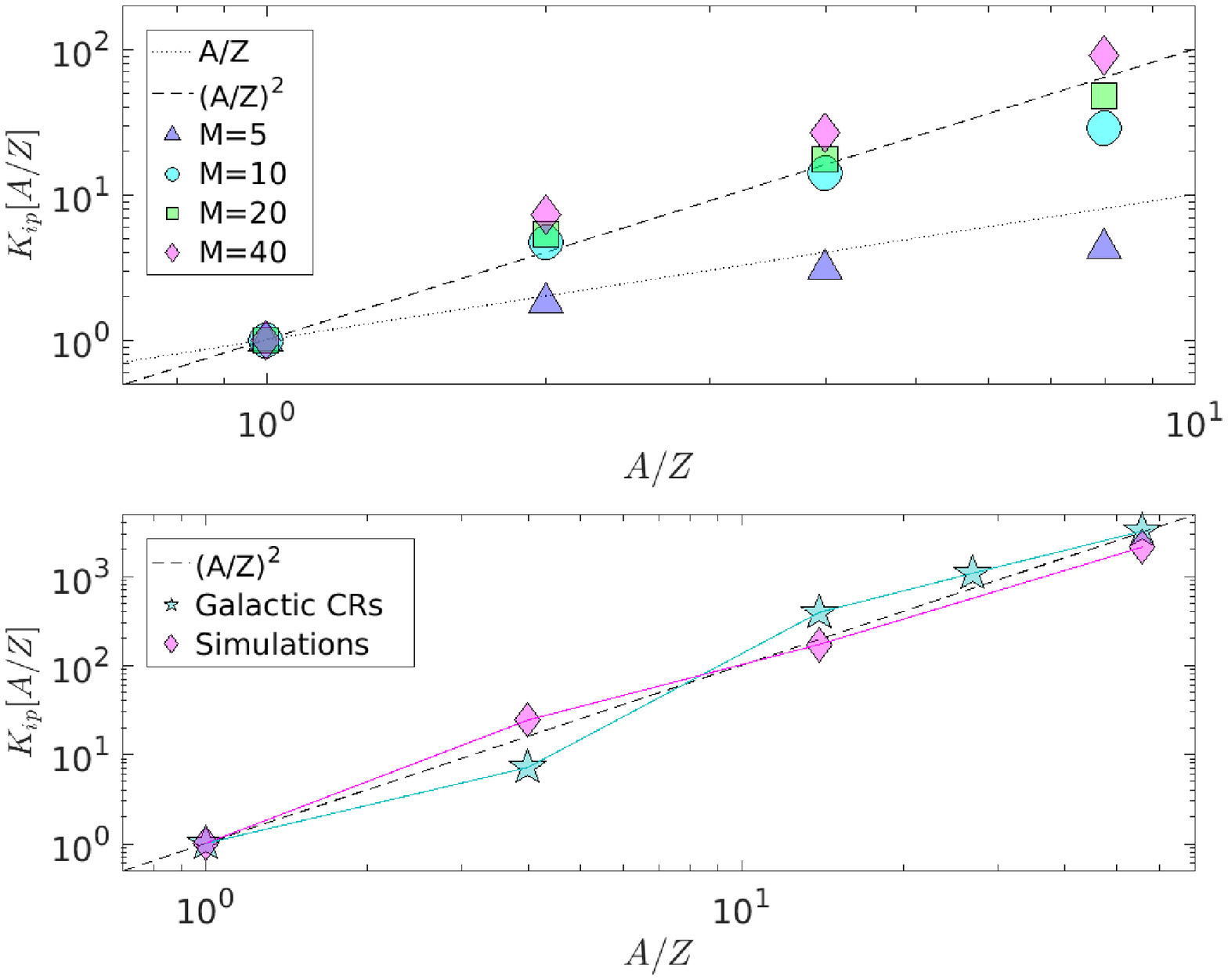}
\caption{Chemical enhancements in GCRs (see Eq.~\ref{eq:kipGCRs}) compared to the ones obtained for a quasi-parallel shock with $M=20$ at $t=10^3\omega_c^{-1}$,  assuming that species are singly ionized. Simulation points are calculated by looking at the spectra of ions reflected in the upstream because Fe ions with $A/Z=56$ have not fully relaxed in the downstream, yet.
The dashed line corresponds to the scaling $\propto (A/Z)^2$ in Fig.~\ref{fig:eta}}
\label{fig:GCRs}
\end{center}
\end{figure}

\emph{Chemical enhancements.---}
The high-$M$ case is relevant, e.g., for SNR shocks propagating into the warm interstellar medium (ISM), where atoms are typically singly ionized.
Ions that are injected into DSA will then be stripped of their electrons while being accelerated up to $\sim$PV rigidities \citep{morlino09,crspectrum}. 
In the popular scenario in which GCRs are produced at SNR shocks via DSA \citep{icrc15}, we can compare our findings with the chemical enrichment measured in GCRs \citep{mde97,nuclei}. In order to compare observations at Earth and shock injection simulations, 
we take the observed GCR flux ratios at 1 TeV, $\phi_i(E)$ \cite[e.g., table 1 in ref.][]{nuclei}, weigh them with the fiducial solar abundances, $\chi_i$ \citep{lodders03}, and write the enhancement at a given $E$ as $\kip Z_i^{1-\gamma}$ (see Eq.~\ref{eq:kip}).
We also account for the rigidity-dependent residence time in the Galaxy $\propto(E/Z)^{-\delta}$, with $\delta\simeq 1/3$ above a few GV \citep{ba11a}, and extrapolate the enhancements down to the non-relativistic injection energies.
Such an extrapolation introduces an additional factor $A_i^{-1/2}$, because DSA spectra are power laws in momentum and hence energy spectra flatten by $E^{1/2}$ at $\sim A_i$GeV.
Finally, we obtain that ion injection into DSA must be enhanced at SNR shocks according to
\begin{equation}\label{eq:kipGCRs}
\kip^{\rm GCRs}=\left.\frac{\phi_i}{\chi_i\phi_p}\right\vert_{\rm TeV}\frac{Z_i^{\gamma-1-\delta}}{A_i^{1/2}}\simeq 
\left.\frac{\phi_i}{\chi_i\phi_p}\right\vert_{\rm TeV}\frac{Z_i^{1/6}}{A_i^{1/2}}
\end{equation}
in order to explain the  abundances observed in GCRs.

We consider a strong quasi-parallel shock with $M=20$ and singly-ionized He, CNO, and Fe atoms with effective $A/Z=\{4,14,56\}$ and calculate $\kip$ in the upstream, since  at $t=10^3\omega_c^{-1}$ ions $A/Z\gtrsim 14$ have already been over-injected but have not yet developed the universal downstream DSA spectrum.
The enhancements found in simulations and those in GCR data (Eq.~\ref{eq:kipGCRs}) are compared in Fig.~\ref{fig:GCRs}: the scaling $\kip\simeq (A/Z)^2$ found for strong shocks provides a very good fit, with singly-ionized He, CNO, and Fe particles enhanced by a factor of about ten, hundred, and a few thousand, respectively.
It is remarkable that such a Fe enhancement requires a very large fraction $\eta_{Fe}\lesssim 50\%$ of the pre-shock Fe ions to enter DSA; 
this may have implications for the overall ISM chemical composition, since regions processed by shocks may become depleted in heavy elements.
Nevertheless, in the ISM many elements are typically trapped in molecules (C,O) and dust grains (Fe, refractory elements), so that fragmentation and sputtering may represent crucial steps in the injection of heavy elements \citep{mde97}.
Our results suggest that dust grains with very large $A/Z\gg1$ should also have no problem of being efficiently energized via DSA-like processes, thereby sputtering pre-energized ions that can be easily injected \citep[][]{edm97,slavin+15}.

In the low-$M$ regime relevant to heliospheric shocks, our simulations show that DSA can account for enhancements by factors of a few to ten, which are commonly observed in SEP events \citep[e.g.,][and references therein]{mason+99,desai+16a}.
However, chemical enhancements in SEPs may be time dependent \citep[e.g.,][]{reames15} and greatly vary from event to event;
in addition to shock strength and inclination, they seem to depend on the presence of pre-existing magnetic turbulence and energetic seed particles (produced, e.g., in solar flares) \citep{tylka+05}, which makes it nontrivial to compare individual SEP events with our simulations where ion injection only occurs from the thermal pool.

\begin{figure*}
\begin{center}
\includegraphics[trim=0px 00px 3px 0px, clip=true, width=0.49\textwidth]{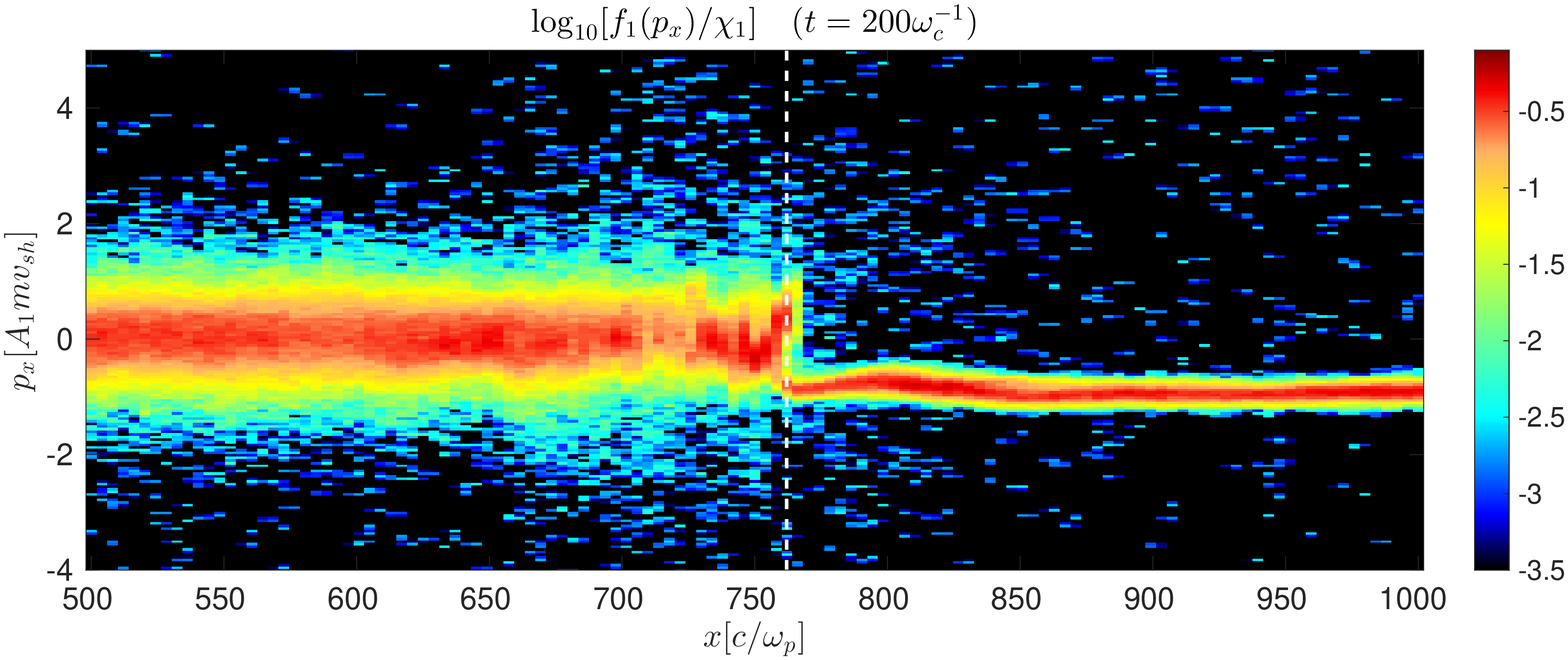}
\includegraphics[trim=0px 0px 3px 0px, clip=true, width=0.49\textwidth]{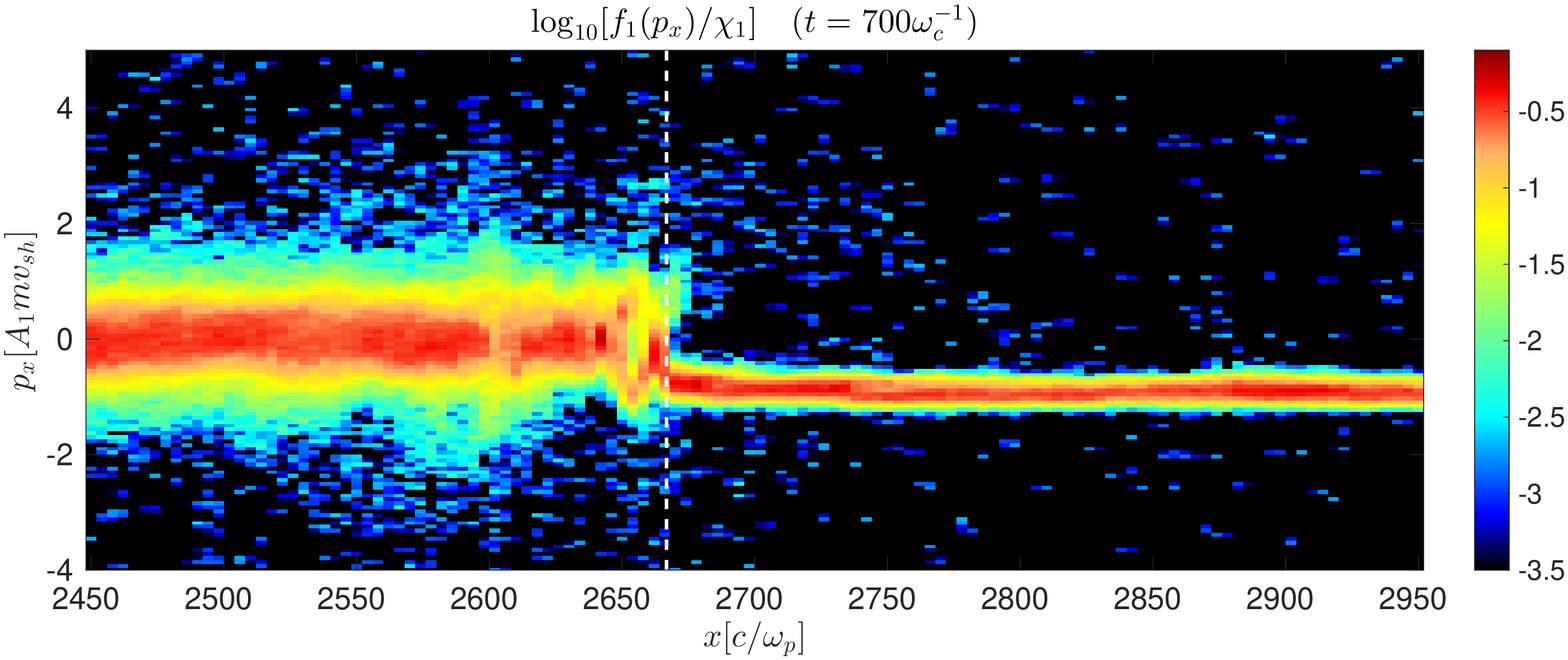}
\includegraphics[trim=0px 0px 3px 0px, clip=true, width=0.49\textwidth]{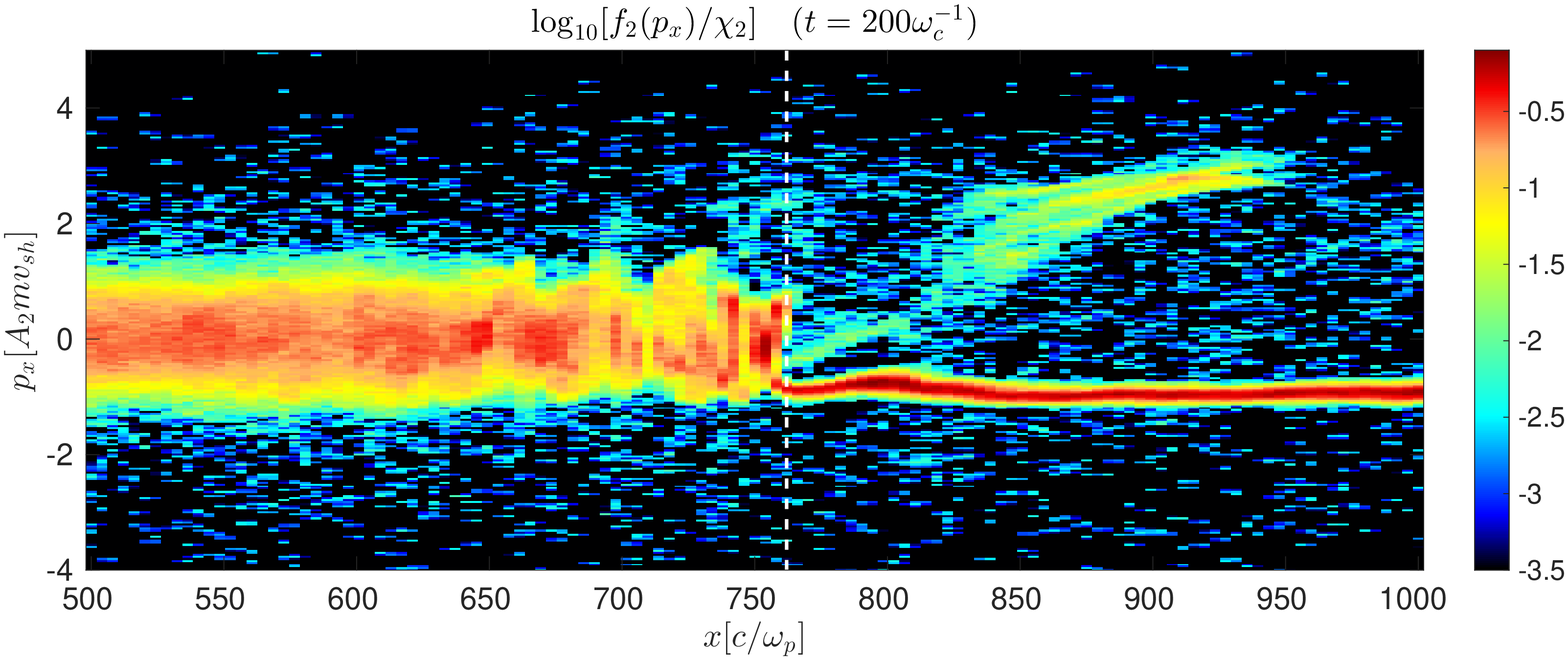}
\includegraphics[trim=0px 00px 3px 0px, clip=true, width=0.49\textwidth]{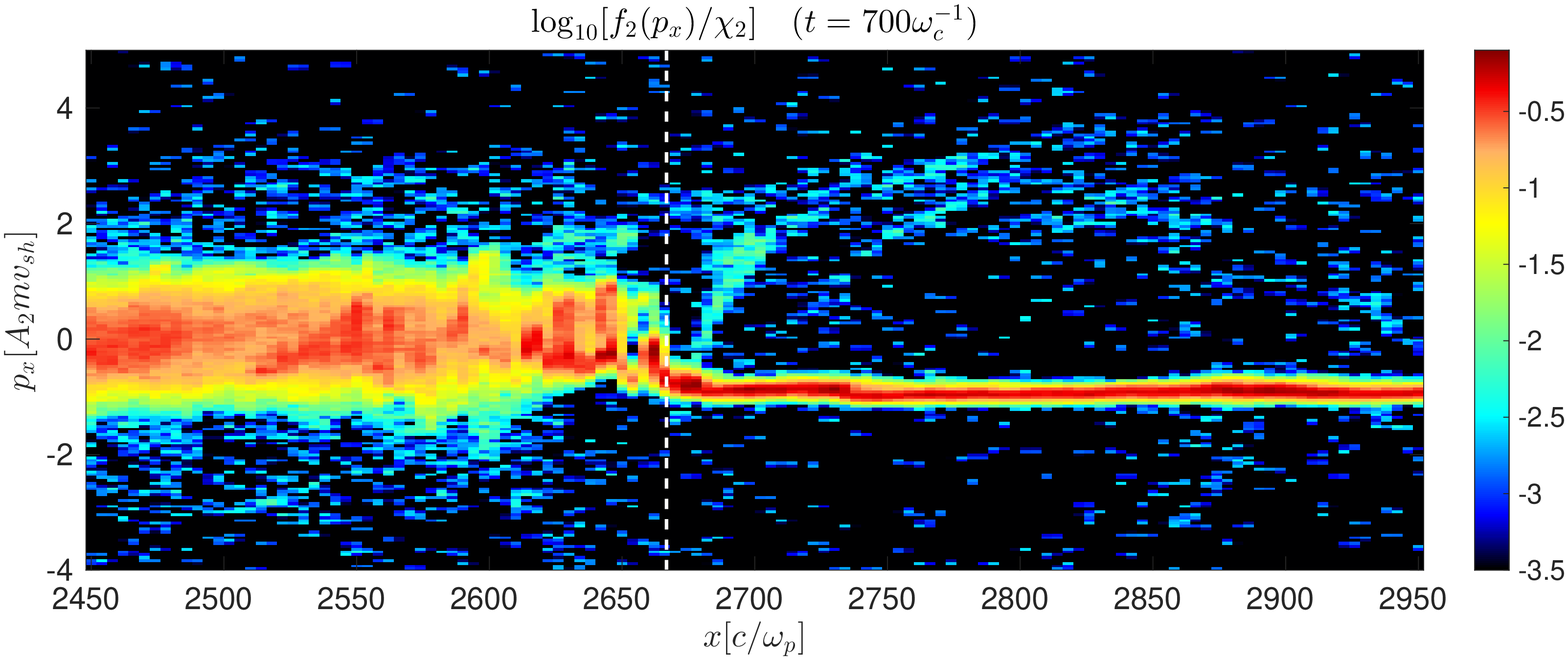}
\includegraphics[trim=0px 0px 3px 0px, clip=true, width=0.49\textwidth]{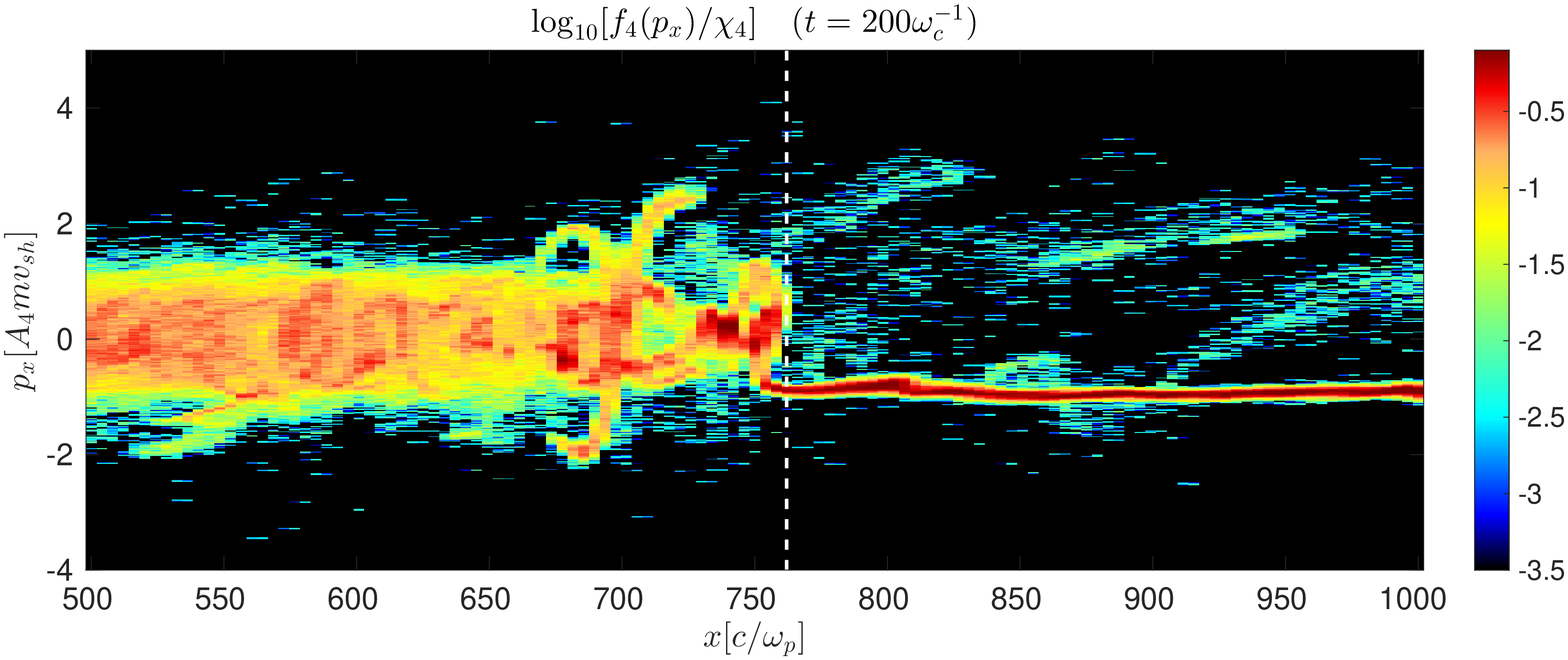}
\includegraphics[trim=0px 00px 3px 0px, clip=true, width=0.49\textwidth]{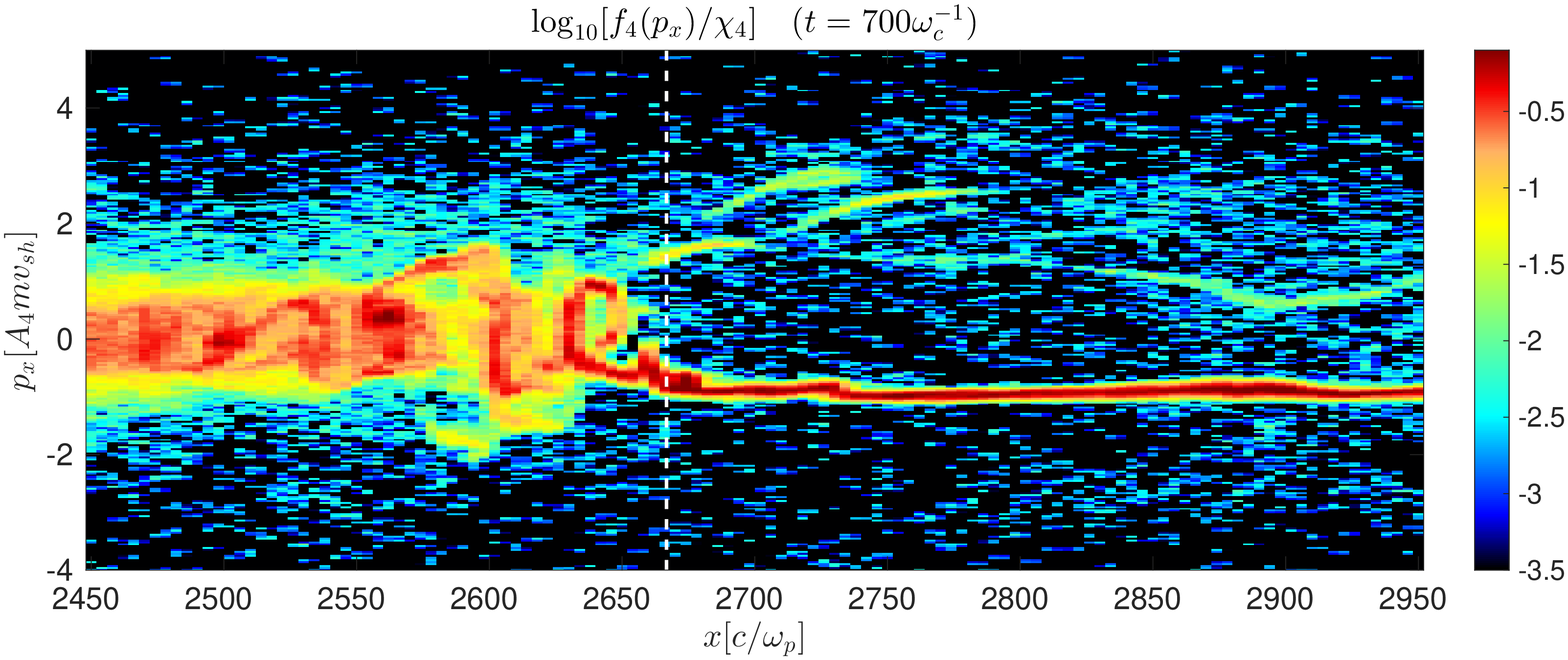}
\includegraphics[trim=0px 00px 3px 0px, clip=true, width=0.49\textwidth]{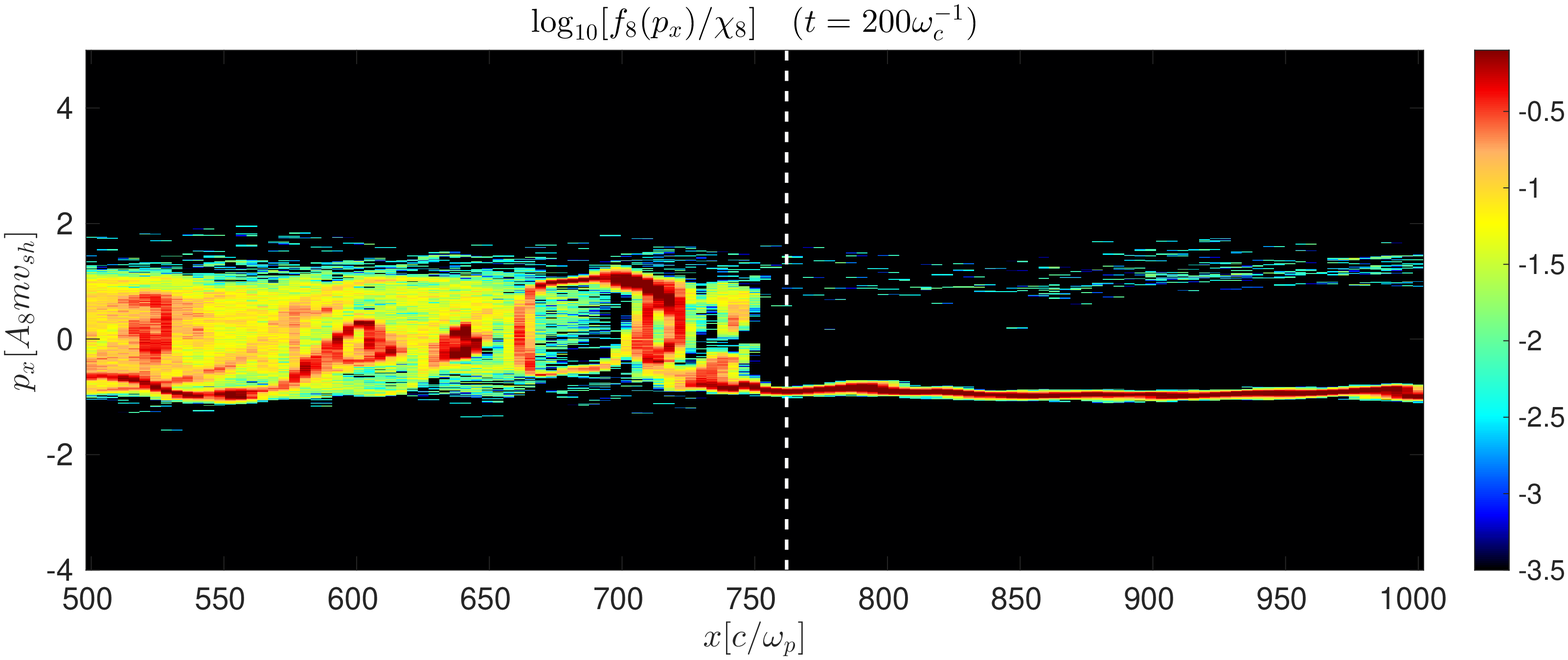}
\includegraphics[trim=0px 0px 3px 0px, clip=true, width=0.49\textwidth]{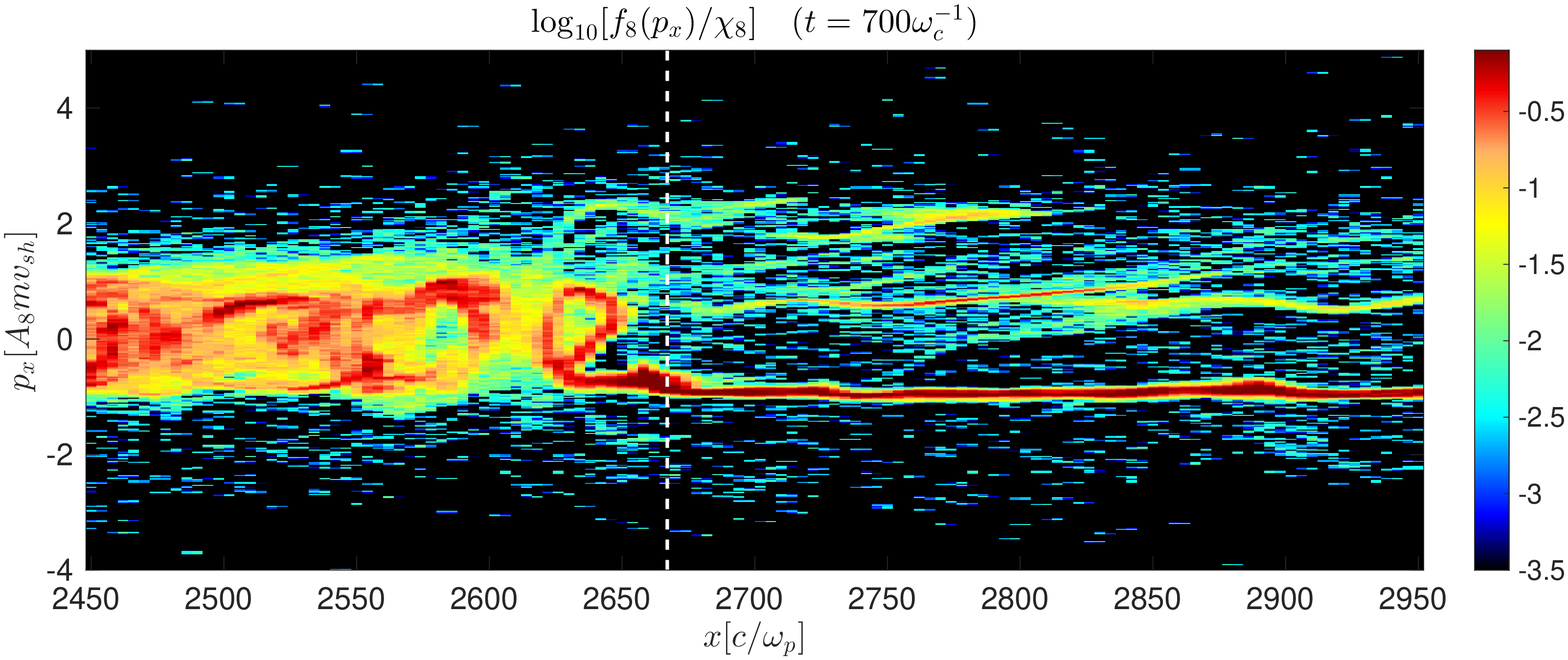}
\caption{$x-p_x$ phase space at $t=\{200,700\}\omega_c^{-1}$ (left to right) for our benchmark case; 
top to bottom panels correspond to species with $A/Z=\{1,2,4,8\}$, respectively.
Ions with larger $A/Z$ isotropize further downstream of the shock marked with dashed lines. 
At $t=200\omega_c^{-1}$ species with $A/Z=\{1,2\}$ already comprise diffusing energetic ($|p|\gtrsim Amv_{sh}$) particles; injection of ions with $A/Z=4$ has just started; 
ions with $A/Z=8$ are still anisotropic in the downstream and not injected yet.
At $t=700\omega_c^{-1}$, instead, all species exhibit typical DSA spectra; the color code shows how the injected fraction increases with $A/Z$.}
\label{fig:px}
\end{center}
\end{figure*}

%%%%%%%%%%%%%%%%%%%%%%%%%%%%%
\emph{Dependence on shock inclination.---}  
Oblique shocks with $\th\gtrsim 50\deg$ cannot inject thermal protons and drive self-generated magnetic turbulence \cite{DSA, injection}.
We find that in such shocks ions with large $A/Z$ do thermalize at a temperature $\propto A$, but progressively further in the downstream with respect to the quasi-parallel case.
Neither protons nor heavier ions are injected into DSA or develop a non-thermal tail, which confirms that having a large gyroradius ($\propto A/Z$) is not a \emph{sufficient} condition for being injected into DSA.

%We have also run simulations with different plasma $\beta\simeq M_A^2/M_s^2$ between 0.1 and 10, finding no appreciable dependence of the ion enhancements on such a parameter as long as both $M_A$ and $M_s$ are large. 
%As usual, the shock dynamics is controlled by the magneto-sonic Mach number $\sim v_{sh}/\sqrt{v_A^2+c_s^2}$, while the amount of magnetic field amplification depends mainly on $M_A$ \citep{MFA}.
%However, since the current instabilities driven by energetic particles have a smaller growth rate in hot plasmas \citep[e.g.][]{reville+08a}, increasing $\beta$ reduces the amplitude of the self-generated fields and eventually the ion enhancements for $M_A\lesssim10$.

\emph{The injection mechanism.---} Ion injection occurs in a qualitatively different way than proton injection, which is due to specular reflection off the time-dependent potential barrier at the shock and energization via shock-drift acceleration \cite{injection}.
Unlike protons, heavy ions are not halted by the shock barrier and always penetrate in the downstream for at least one gyroradius ($\sim MA/Z c/\omega_p$); here, their distribution tends to become more isotropic due to the presence of rapidly-varying fields, an analog of the violent relaxation in stellar dynamics.
If isotropization is rapid enough with respect to advection, there arises a population of backstreaming ions that can overrun the shock barrier, which is ``tuned'' for preventing downstream thermal \emph{protons} from returning upstream.
The fraction of injected heavy ions is thus controlled by how rapid isotropization is, which depends on $A/Z$ and on the strength of the magnetic turbulence in the shock layer.

Fig.~\ref{fig:px} shows the $x-p_x$ phase spaces for our benchmark run; we consider ions with $Z=1$ and $A=\{1,2,4,8\}$ at times $t=\{200,700\} \omega_c^{-1}$.
We see that, while protons are promptly isotropized behind the shock (dashed vertical lines), ions with larger $A$ tend to retain their anisotropy further in the downstream. 
At early times (left column), protons show the characteristic non-thermal, isotropic population of particles diffusing around the shock \citep{DSA};
ions with $A/Z=2$ have also started being injected and accelerated.
Injection may be quite ``bursty'' when patches of quasi-parallel magnetic field are advected through the shock \citep{injection}, resulting  in coherent batches of particles protruding back into the upstream (as for $A/Z=2$ at $t=200 \omega_c^{-1}$ in Fig.~\ref{fig:px}).
Ions with $A/Z=4$ have just started overrunning the shock, but there are only few particles with $p_x<0$ in the upstream, implying that DSA has not yet been established.
Finally, ions with $A/Z=8$ isotropize too far downstream to overrun the shock barrier and are not injected yet.

At later times ($t=700 \omega_c^{-1}$, right column in Fig.~\ref{fig:px}), instead, all of the species show the typical DSA spectrum comprising non-thermal particles that diffuse on both sides of the shock.
From the color code it is also possible to see how the fraction of particles that leak back into the upstream is larger for heavier ions.

Proton injection is controlled by the quasi-periodic reformation of the shock barrier \citep{injection}; 
instead, injection of heavier ions relies on rapid electromagnetic fluctuations larger than those induced by the local shock reformation and happens at later times for heavier species, and always after the onset of non-linear turbulence \footnote{The expected dependence $E_{max,i}(t)\propto Z$ is achieved only when the acceleration time becomes much larger than the injection time, which may explain why in some heliospheric shocks Fe spectra roll over at significantly lower energies compared with O \citep{desai+16a}.}.

A more quantitative characterization of the trajectories of the ions that get injected into DSA is beyond the scope of this Letter. 
Note, however, that the behavior reported here is \emph{not} equivalent to the so-called \emph{thermal leakage} scenario for particle injection \citep[e.g.,][]{edmiston+82,malkov98,bgv05}, in that the injected ions are not those in the tail of the Maxwellian (strictly speaking, they have not yet thermalized).
The global shock structure is always controlled by species with the most inertia, so that density and fields jump within one gyroradius of thermal protons. 
The isotropization length for heavy ions is effectively larger than for protons, but injection is controlled by how rapidly they can be isotropized (i.e., reverse their $p_x$), which depends on the local electromagnetic fluctuations and not on the ion energy.

We conclude that, while the injected protons are reflected by the shock barrier and need to be pre-energized via few cycles of shock-drift acceleration \citep{injection}, heavy ions reflect off post-shock magnetic irregularities. 
The enhancement in ions with $A/Z\gg1$ is then due to the fact that they are not affected by the proton-regulated shock barrier (their kinetic energy being much larger than the barrier potential), so that they do not experience shock-drift acceleration but rather start diffusing right away. 
Ions with $A/Z\gtrsim 1$ exhibit intermediate properties between protons and heavier ions because their probability of being reflected or transmitted at the shock barrier depends on the actual angle between their momentum and the shock normal: ions with velocity mainly along the shock surface are more proton-like because they can be reflected by the shock barrier \citep{injection}.

\emph{Conclusions.---}
We have presented the first ab-initio calculation of ion DSA at non-relativistic shocks, finding that species with large $A/Z$ show enhanced non-thermal tails with respect to protons, in quantitative agreement with the chemical abundances observed in GCRs.
In forthcoming publications we will discuss the implications of these findings also for what concerns the discrepant hardening of non-H species in GCRs \citep[e.g.,][]{cream10} and for the role of accelerated He in SNR shocks \citep{nuclei}.

\begin{acknowledgments}
This research was supported by NASA (grant NNX17AG30G to DC), NSF (grant AST-1517638 to AS), and Simons Foundation (grant 267233 to AS). Simulations were performed on computational resources provided by the Princeton High-Performance Computing Center, the University of Chicago Research Computing Center, and XSEDE TACC (TG-–AST100035).
\end{acknowledgments}

\bibliography{Total}

%  aspect ratio (width / height); word count [(150 / aspect ratio) + 20 words]
%  aspect ratio (width / height); word count [(600 / aspect ratio) + 40 words]
% fig 1= 2.3            => 300
% fig 2= 7/6            => 149
% fig 3= 7/3 =          => 84
% fig 4= 680/570 =      => 543   TOT= 1,124 words in figures
% 3 eqs                 => 48    TOT= 48;
%words in eqs + fig                 = 1,172
% wc   = 26442644-165(preamble)   => 20    TOT= 2,479 (target: 2,578)
%                                MAX= 3750
\end{document}